\documentclass{article}
\usepackage[T1]{fontenc}

\usepackage{spconf,amsmath,graphicx,hyperref}

\usepackage{amsmath,amssymb,amsfonts}
\usepackage{algorithmic}
\usepackage{graphicx}
\usepackage{textcomp}
\usepackage{stfloats}
\usepackage{url}
\usepackage{verbatim}
\usepackage{graphicx}
\usepackage{cite}
\usepackage{multirow}
\usepackage{lscape,array}
\usepackage{booktabs}
\usepackage{pifont}
\usepackage{ragged2e}
\usepackage{enumitem}
 \usepackage{xcolor}
\usepackage[table]{xcolor}
\usepackage{hyperref}
\usepackage{balance}

\title{\textbf{\Large BUT Systems for WildSpoof Challenge: SASV in the Wild}}

\name{
Junyi Peng$^1$, Jin Li$^{1,3}$, Johan Rohdin$^1$, Lin Zhang$^2$, Miroslav Hlav\v{áč}ek$^1$, Old\v{r}ich Plchot$^1$
}

\address{
        $^1$Speech@FIT, Brno University of Technology, Czechia
        $^2$Johns Hopkins University, USA \\
        $^3$Department of EEE, The Hong Kong Polytechnic University, Hong Kong SAR
    }

\begin{document}
\maketitle

\begin{abstract}
This paper presents the BUT submission to the WildSpoof Challenge, focusing on the Spoofing-robust Automatic Speaker Verification (SASV) track. We propose a SASV framework designed to bridge the gap between general audio understanding and specialized speech analysis. Our subsystem integrates diverse Self-Supervised Learning front-ends ranging from general audio models (e.g., Dasheng) to speech-specific encoders (e.g., WavLM). These representations are aggregated via a lightweight Multi-Head Factorized Attention back-end for corresponding subtasks. Furthermore, we introduce a feature domain augmentation strategy based on Distribution Uncertainty to explicitly model and mitigate the domain shift caused by unseen neural vocoders and recording environments. By fusing these robust CM scores with state-of-the-art ASV systems, our approach achieves superior minimization of the a-DCFs and EERs.

\end{abstract}
\begin{keywords}
Self-supervised learning, speaker verification, anti-spoofing, fine-tuning
\end{keywords}
\section{Introduction}
Although previous challenges like ASVspoof~\cite{wang2020asvspoof,rohdin2024but} have advanced the field of Spoofing-robust ASV (SASV), they often rely on clean studio-recorded bona fide speech, creating a mismatch with real-world deployment scenarios where noise and reverberation are ubiquitous. 
The newly introduced \textbf{WildSpoof Challenge}~\cite{wu2025wildspoof} and the \textbf{SpoofCeleb} dataset~\cite{jung2025spoofceleb} address this gap by deriving bona fide speech from VoxCeleb1~\cite{nagrani2017voxceleb} and generating spoofing attacks using TTS systems trained on the same noisy data.

The core task of WildSpoof is SASV, which requires a system to accept only bona fide target trials while rejecting both zero-effort impostors (non-target) and spoofing attacks. 
This presents a core challenge: the system must be robust to the ``generation-recognition trade-off''~\cite{jung2025spoofceleb}, effectively distinguishing forensic artifacts in noisy conditions where traditional detection cues might be masked.

In this paper, we present the \textbf{BUT} submission to the WildSpoof SASV track. 
Our approach focuses on maximizing the representation power of diverse Self-Supervised Learning (SSL) models for both the anti-spoofing countermeasure (CM) and automatic speaker verification (ASV) systems. 
We hypothesize that large-scale pre-trained models encode rich acoustic information that, when properly aggregated, is resilient to environmental noise~\cite{peng2025mhfa}. 
Crucially, we propose a flexible framework designed to be compatible with both general audio SSLs (e.g., Dasheng~\cite{dinkel2024scaling}) and speech-specific SSLs (e.g., WavLM\cite{chen2022wavlm}, W2V2-BERT\cite{barrault2023seamless}). This allows us to leverage the broad acoustic understanding of audio models alongside the specialized phonetic features of speech models. Moreover, we utilize a \textbf{Multi-Head Factorized Attention (MHFA)} backend~\cite{peng2022attention} augmented with a \textbf{Distribution Uncertainty (DSU}) module~\cite{peng2025but} to simulate unseen domain shifts.

Our main contributions are summarized as follows:
\begin{itemize}[nosep]
    \item We demonstrate that general audio SSL models (e.g., Dasheng) provide complementary robustness to speech-specific models when applied to noisy, in-the-wild deepfake detection within our proposed framework.
    \item We integrate DSU-based feature augmentation into the MHFA backend, significantly improving performance on unseen attacks in the SpoofCeleb evaluation set.
\end{itemize}





\section{Proposed Method}
\label{sec:method}
\subsection{Data and Framework}
The proposed system is developed within the \textbf{WeDefense}~\cite{zhang2026wedefense} framework\footnote{\url{https://github.com/zlin0/wedefense}}, an open-source toolkit specifically designed for defending against fake audio. We utilize the official \textbf{SpoofCeleb} dataset as the primary source for both training and developing our CM models. Furthermore, to enhance the speaker verification component, we incorporate the \textbf{VoxCeleb2-dev} dataset for training our SV systems.

\subsection{Hierarchical SSL Feature Extraction}
SSL models, such as WavLM~\cite{chen2022wavlm}, HuBERT~\cite{hsu2021hubert}, and Dasheng~\cite{dinkel2024scaling}, encode rich acoustic information across their layers. Lower layers typically capture raw spectral details, while upper layers encode more semantic or structural information~\cite{peng2024probing,10887574}. For the task of deepfake detection, relying on the last layer is often sub-optimal, as forensic artifacts introduced by neural vocoders (e.g., phase discontinuities or metallic buzzing) are often best preserved in intermediate representations.

Therefore, instead of using only the final-layer hidden state, we employ MHFA, a lightweight backend that learns layer-wise attention weights and computes a weighted sum of all $L$ transformer layers. Let $X \in \mathbb{R}^{L \times T \times D}$ be the output features from all layers of the SSL encoder, where $T$ is the number of frames and $D$ is the feature dimension. We learn layer-specific weights to aggregate these features dynamically, allowing the backend to focus on the most discriminative level of abstraction.


\subsection{Multi-Head Factorized Attention (MHFA)}
To effectively aggregate the temporal frame-level features into a global utterance-level embedding, we employ the \textbf{MHFA} mechanism~\cite{peng2022attention}. Unlike standard attention, which uses a single linear projection, MHFA factorizes the aggregation process into two independent streams: a \textit{Key} stream ($K$) and a \textit{Value} stream ($V$).

Specifically, we define two separate sets of learnable layer weights, $w^k \in \mathbb{R}^L$ and $w^v \in \mathbb{R}^L$. These weights are normalized via softmax to compute the weighted sum of the SSL layer outputs $Z_l$:
\begin{equation}
    K_{feat} = \sum_{l=1}^{L} \text{softmax}(w^k_l) Z_l, \quad V_{feat} = \sum_{l=1}^{L} \text{softmax}(w^v_l) Z_l
\end{equation}
These aggregated features are then projected into a lower dimension $D_{cmp}$ using linear layers $W_k$ and $W_v$:
\begin{equation}
    K = K_{feat} W_k, \quad V = V_{feat} W_v
\end{equation}
The attention weights $A$ are computed from the Query stream, while the content to be aggregated comes from the Value stream. This factorization allows the model to learn \textit{where} to look using $K$ independently of \textit{what} to extract using $V$. For $H$ heads, the output is pooled as:
\begin{equation}
    A = \text{softmax}(K W_{att}, \text{dim}=1)
\end{equation}
\begin{equation}
    Embedding = \text{Pooling}(V \odot A)
\end{equation}
Finally, a fully connected layer maps the concatenated head outputs to the final embedding $e$. This embedding is then processed by the corresponding classification head based on the specific task.

\begin{table*}[t]
\centering
\caption{ASV Performance Comparison (EER \%) (spoofed trials are excluded.).
}
\label{tab:asv_results}
\begin{tabular}{lccccc}
\toprule
\textbf{System} & \textbf{Vox1-O} & \textbf{Vox1-E} & \textbf{Vox1-H} & \textbf{SpoofCeleb-Dev (SV)} & \textbf{SpoofCeleb-Eval (SV)} \\
\midrule
ResNet293 & 0.447 & 0.657 & 1.183 & 3.209 & 3.053 \\
WavLM Large + MHFA & 0.516 & 0.583 & 1.179 & 3.273 & 3.510 \\
W2V2-BERT + MHFA & \textbf{0.229} & \textbf{0.354} & \textbf{0.714} & \textbf{2.441} & \textbf{2.528} \\
\bottomrule
\end{tabular}
\end{table*}

\begin{table*}[t]
\centering
\caption{Performance of different CM systems trained on SpoofCeleb. *DSU indicates Distribution Uncertainty augmentation.}
\label{tab:cm_results}
\begin{tabular}{lcccccc}
\toprule
\multirow{2}{*}{\textbf{System}} & \multicolumn{2}{c}{\textbf{SpoofCeleb Dev}} & \multicolumn{2}{c}{\textbf{SpoofCeleb Eval}} & \multicolumn{2}{c}{\textbf{OOD (ASV5 Dev)}} \\
& \textbf{EER(\%)} & \textbf{minDCF} & \textbf{EER(\%)} & \textbf{minDCF} & \textbf{EER(\%)} & \textbf{minDCF} \\
\midrule
\textbf{Speech-Specific SSL} & & & & & & \\
WavLM Base+ & 0.402 & 0.011 & 0.055 & 0.001 & 11.885 & 0.193 \\
\quad + MUSAN/RIR & 0.805 & 0.022 & 0.153 & 0.003 & 7.976 & 0.146 \\
\quad + RawBoost & \textbf{0.103} & \textbf{0.003} & \textbf{0.041} & \textbf{0.001} & 6.970 & 0.158 \\
\midrule
\textbf{General Audio SSL} & & & & & & \\
Mi-Dasheng-base (86M) & 0.239 & 0.006 & 0.123 & 0.003 & 5.164 & 0.137 \\
Mi-Dasheng-0.6B (600M) & 0.176 & 0.004 & 0.050 & 0.001 & 3.122 & 0.089 \\
\quad + DSU & 0.213 & 0.005 & 0.078 & 0.002 & \textbf{1.777} & \textbf{0.051} \\
Mi-Dasheng-1.2B (1200M) & 0.265 & 0.006 & 0.090 & 0.002 & 1.625 & 0.047 \\
\quad + DSU & 0.301 & 0.007 & 0.154 & 0.003 & \textbf{1.193} & \textbf{0.034} \\
\midrule
\textbf{Baseline} & & & & & & \\
ResNet18 & 0.204 & 0.005 & 0.185 & 0.005 & 12.090 & 0.177 \\
\bottomrule
\end{tabular}

\end{table*}

\subsection{MHFA with DSU (Feature Domain Augmentation)}
To tackle the challenge of unseen generators, we integrate a Feature Domain Augmentation strategy directly into the MHFA backend~\cite{li2022uncertainty}, termed \textbf{MHFA-DSU}. This method is based on the concept of Distribution Uncertainty (DSU), which hypothesizes that domain shifts can be simulated by perturbing the feature statistics (mean and variance) of the training data.

We apply DSU specifically to the Value stream ($V_{feat}$) before the linear projection. During training, with a probability $p$, we model the feature statistics as distributions rather than deterministic values. For an input feature map $x$ (corresponding to $V_{feat}$), we compute the instance-level mean and standard deviation across the temporal dimension \cite{li2022uncertainty}. This operation essentially ``jitters'' the global style and channel characteristics of the audio representation while preserving the local content, forcing the network to learn features that are invariant to global statistical shifts caused by different vocoders.

\subsection{Calibration and Fusion}\label{sec:calibration}
To fuse and calibate decisions from ASV and CM, our report discussed two methods:
\begin{itemize}[nosep]
    \item Pre-fusion individual calibration: calibrate ASV and CM scores separately with logistic regression, then fuse (optionally followed by a final calibration). 
    \item Joint calibration/fusion: jointly learn scale and bias for ASV and CM within a single logistic fusion model (as in Eq. 9 of our previous work~\cite{rohdin2024but}).
\end{itemize}

\section{Experiments}

\subsection{Experimental Setup}
Our models were implemented using PyTorch and trained on AMD Instinct MI200 GPUs. The training process was configured with a maximum of 8 epochs. We utilized the AdamW optimizer with $\beta_1=0.9$, $\beta_2=0.999$, and a weight decay of $1.0 \times 10^{-4}$. The batch size was set to 128.

We employed a differential learning rate strategy to fine-tune the SSL front-end and the MHFA back-end effectively. The base learning rate was initialized at $5.0 \times 10^{-4}$ and decayed to a final learning rate of $1.0 \times 10^{-5}$ using a Cosine Annealing scheduler. To prevent catastrophic forgetting of the pre-trained representations, the learning rate for the SSL front-end was scaled by a factor of 0.05 relative to the base learning rate. A warmup period of 2 epochs was applied at the beginning of training.

For the MHFA backend configuration, we set the number of attention heads (\texttt{head\_nb}) to 32, the embedding dimension (\texttt{embed\_dim}) to 256, and the compression dimension (\texttt{compression\_dim}) to 128.

\subsection{Results and Analysis}
\subsubsection{ASV}
We first evaluate the performance of our ASV subsystems. Table \ref{tab:asv_results} summarizes the results of three different ASV models on the standard VoxCeleb test sets (Vox1-O, Vox1-E, Vox1-H) and the SpoofCeleb Development and Evaluation sets. All models were trained on the VoxCeleb2-dev dataset with Additive Angular Margin Softmax (AAM-Softmax) as loss function. And  VoxCeleb2-dev is used for cosine score normalization.

The W2V2+BERT + MHFA model demonstrates superior performance across all evaluation sets, significantly outperforming the ResNet293 baseline and the WavLM Large model. Specifically, on the challenging SpoofCeleb-Dev and SpoofCeleb-Eval sets, which contain in-the-wild noisy speech, W2V2+BERT + MHFA achieves the lowest EERs of 2.440\% and 2.250\%, respectively. This highlights the robustness of the MHFA backend combined with rich self-supervised representations in handling diverse acoustic conditions.

\subsubsection{Anti-Spoofing}
We evaluate various CM systems utilizing different SSL backbones and augmentation strategies, including WavLM Base+ and Mi-Dasheng models.
For all of them, we utilize a standard Binary Cross-Entropy (BCE) loss to distinguish between \textit{bonafide} and \textit{spoof} classes.
Table \ref{tab:cm_results} presents the results on SpoofCeleb Dev/Eval sets and an out-of-domain (OOD) set (ASVSpoof5 Dev).

\textbf{General Audio vs. Speech SSLs:} General audio models (Dasheng) consistently outperform speech-specific models (WavLM) and the ResNet18 baseline on the OOD dataset. For instance, Mi-Dasheng-0.6B achieves 3.122\% EER on ASV5 Dev compared to 11.885\% for WavLM, demonstrating superior generalization. We further observe that Mi-Dasheng-0.6B slightly outperforms Mi-Dasheng-1.2B, suggesting that larger model capacity does not necessarily yield better performance, and selecting an appropriately sized backbone for the data and task can be more effective.

\textbf{Effectiveness of DSU:} Although applying DSU augmentation doesn't show improvement on the in-domain data, it shows significantly improved robustness on OOD data. For Mi-Dasheng-0.6B, DSU reduces the OOD EER from 3.122\% to 1.777\%. Similarly, for the 1.2B model, DSU improves OOD EER from 1.625\% to 1.193\%, validating its efficacy in handling unseen domains.

\subsection{Calibration and Fusion}

\begin{table*}[!ht]
\caption{Results of fusion and calibration on ASV and CM systems. (Only spoofceleb data are used for calibration)}
\label{tab:asv_cm_fusion_cali}
\centering
\scalebox{0.95}{
\begin{tabular}{llllll}
\toprule
\multicolumn{2}{c}{\textbf{CM}}                   & \multicolumn{2}{c}{\textbf{ASV}}           & \textbf{SASV}            & \multicolumn{1}{c}{\textbf{Spoofceleb dev.}} \\
\textbf{CM model}         & \textbf{calibrate on} & \textbf{ASV model} & \textbf{calibrate on} & \textbf{fusion/cali. on} & \multicolumn{1}{l}{\textbf{a-DCF}}           \\
\midrule
Dasheng 0.6B + MHFA + DSU & dev + eval & W2V2-BERT + MHFA   & dev + eval & dev+eval      & 0.02747                                      \\
Dasheng 0.6B + MHFA + DSU & no                    & W2V2-BERT + MHFA   & dev + eval & dev+eval      & 0.02747                                      \\
Dasheng 0.6B + MHFA       & dev + eval & W2V2-BERT + MHFA   & dev + eval & dev+eval      & 0.02695                                     \\
\bottomrule
\end{tabular}
}
\end{table*}

Table~\ref{tab:asv_cm_fusion_cali} reports results for the two calibration strategies described in Section\ref{sec:calibration}. We observe that calibrating component scores before fusion versus calibrating the score joinlty during fusion yields similar performance. This suggests that separately calibrating the CM system is unnecessary if a joint calibration is applied. 
However, it remains unexplored whether joint optimization can be numerically challenging without pre individual calibration before fusion for more difficult sets or for the ASV system which, due to the properties of cosine scoring, has more constrained raw scores. 

\section{Conclusion and Discussion}
This report describes the BUT systems for the WildSpoof Challenge 2025 SASV track. For ASV, we compare ResNet with popular SSL backbones, and for CM, we use general‑audio SSL models (Dasheng) with DSU‑based feature augmentation and a lightweight MHFA backend. For fusion/calibration, we compare pre‑fusion individual calibration and joint calibration/fusion. Results show that general‑audio models outperform speech‑specific models and the ResNet18 baseline on OOD data, DSU improves generalization, and pre‑ vs post‑fusion calibration performs similarly. 

{\footnotesize
\bibliographystyle{IEEEtran}
\bibliography{mybib}
}
\end{document}